\documentclass[aip,apl,reprint,a4paper,floatfix,amsfonts,amsmath,amssymb,citeautoscript,groupedaddress,twocolumn]{revtex4}

\usepackage{mathptmx}
\usepackage[scaled=0.9]{helvet}
\usepackage[utf8]{inputenc}
\usepackage{graphicx}
\usepackage[pdftex,dvipsnames]{xcolor}
\usepackage[pdftex]{hyperref}
\hypersetup{
    pdftitle={Counterintuitive strain distribution in axial (In,Ga)N/GaN nanowires},
    colorlinks,
    citecolor=blue
    }
\usepackage{upgreek}
\usepackage{bm}
\usepackage{textcomp}
\usepackage[
,textwidth=17.5cm
,textheight=23.5cm
,verbose
,dvips
]{geometry}
\usepackage{xspace}
\usepackage{xargs}
\usepackage[colorinlistoftodos,linecolor=Plum,backgroundcolor=Plum!25,bordercolor=Plum,prependcaption,textsize=footnotesize,textwidth=1.4cm]{todonotes}

\newcommand{\abs}[1]{{\left| #1 \right|}}
\newcommand{\ii}{\mathrm{i}}

\newcommand{\exx}{$\upvarepsilon_\mathrm{xx}$\xspace}
\newcommand{\eyy}{$\upvarepsilon_\mathrm{yy}$\xspace}
\newcommand{\ezz}{$\upvarepsilon_\mathrm{zz}$\xspace}

\newcommand{\eyz}{$\upvarepsilon_\mathrm{yz}$\xspace}
\newcommand{\ezx}{$\upvarepsilon_\mathrm{zx}$\xspace}

\newcommand{\qx}{q$_\mathrm{x}$\xspace}
\newcommand{\qy}{q$_\mathrm{y}$\xspace}
\newcommand{\qz}{q$_\mathrm{z}$\xspace}

\newcommand{\ingan}{In$_{0.3}$Ga$_{0.7}$N\xspace}
\newcommand{\ingagan}{In$_{0.3}$Ga$_{0.7}$N/GaN\xspace}

\begin{document}

\title{Counterintuitive strain distribution in axial (In,Ga)N/GaN nanowires}
\author{Thilo Krause}
\email{krause@pdi-berlin.de}
\author{Michael Hanke}
\author{Oliver Brandt}
\author{Achim Trampert}
\affiliation{Paul-Drude Institut für Festkörperelektronik, Hausvogteiplatz 5--7, 10117 Berlin, Germany}

\begin{abstract}
We study the three-dimensional deformation field induced by an axial (In,Ga)N segment in a GaN nanowire. Using the finite element method within the framework of linear elasticity theory, we study the dependence of the strain field on the ratio of segment length and nanowire radius. Contrary to intuition, the out-of-plane-component \ezz of the elastic strain tensor is found to assume large negative values for a length-to-radius ratio close to one. We show that this unexpected effect is a direct consequence of the deformation of the nanowire at the free sidewalls and the associated large shear strain components. Simulated reciprocal space maps of a single (In,Ga)N/GaN nanowire demonstrate that nanofocus x-ray diffraction is a suitable technique to assess this peculiar strain state experimentally.
\end{abstract}

\pacs{
87.10.Kn
, 81.05.Ea
, 81.07.Gf
, 61.10.Nz
}

\maketitle

(In,Ga)N/GaN heterostructures are an integral part of light emitting devices used for full-color displays and solid-state lighting. For conventional planar heterostructures, the In content in the (In,Ga)N quantum well is restricted due to the formation of dislocations beyond a certain critical stress\cite{Matthews1974,Matthews1975,Parker1999,Pereira2002,Dobrovolskas2013}. Axial (In,Ga)N/GaN nanowire (NW) heterostructures are promising alternatives since their high aspect ratio and high surface-to-volume ratio are expected to facilitate the incorporation of higher amounts of In without resulting in plastic deformation\cite{Ertekin2005,Hersee2011}. In fact, a NW can effectively release strain elastically close to its free surface, and thereby accommodate a higher lattice mismatch compared to a planar layer\cite{Glas2006,Ye2009}. The prospect to incorporate a high amount of In into a GaN matrix offers the intriguing possibility to tune the emitted wavelength from the near-ultraviolet to the near-infrared region\cite{Kuykendall2007}. The potential advantages of NWs for light emitting applications have encouraged many investigations of their growth\cite{Johansson2011,Li2012}, related phenomena such as phase segregation\cite{Segura-Ruiz2011,Segura-Ruiz2014} as well as their optical properties\cite{Limbach2012,Marquardt2013,Jackson2014}.

The elastic strain relaxation in axial NW heterostructures depends on both the length of the strained segment and the radius of the NW, which therefore represent additional degrees of freedom to tune the emission wavelength\cite{Kaganer2012,Wolz2013}. In contrast to planar structures, however, the strain distribution in a NW heterostructure is inherently three-dimensional and more complex than commonly assumed.

In this letter, we investigate the strain field in axial (In,Ga)N/GaN NW heterostructures using the finite element method. In particular, we discuss the dependence of the out-of-plane component of the elastic strain tensor, \ezz, on the length of the (In,Ga)N segment and the NW radius. We show that for certain length-to-radius ratios, \ezz may assume large negative values. Finally, we suggest that x-ray diffraction experiments on single (In,Ga)N/GaN NWs employing nanofocus synchrotron radiation are a viable means to access the strain state in these nanostructures experimentally.

\begin{figure}
\includegraphics[width=0.8\columnwidth]{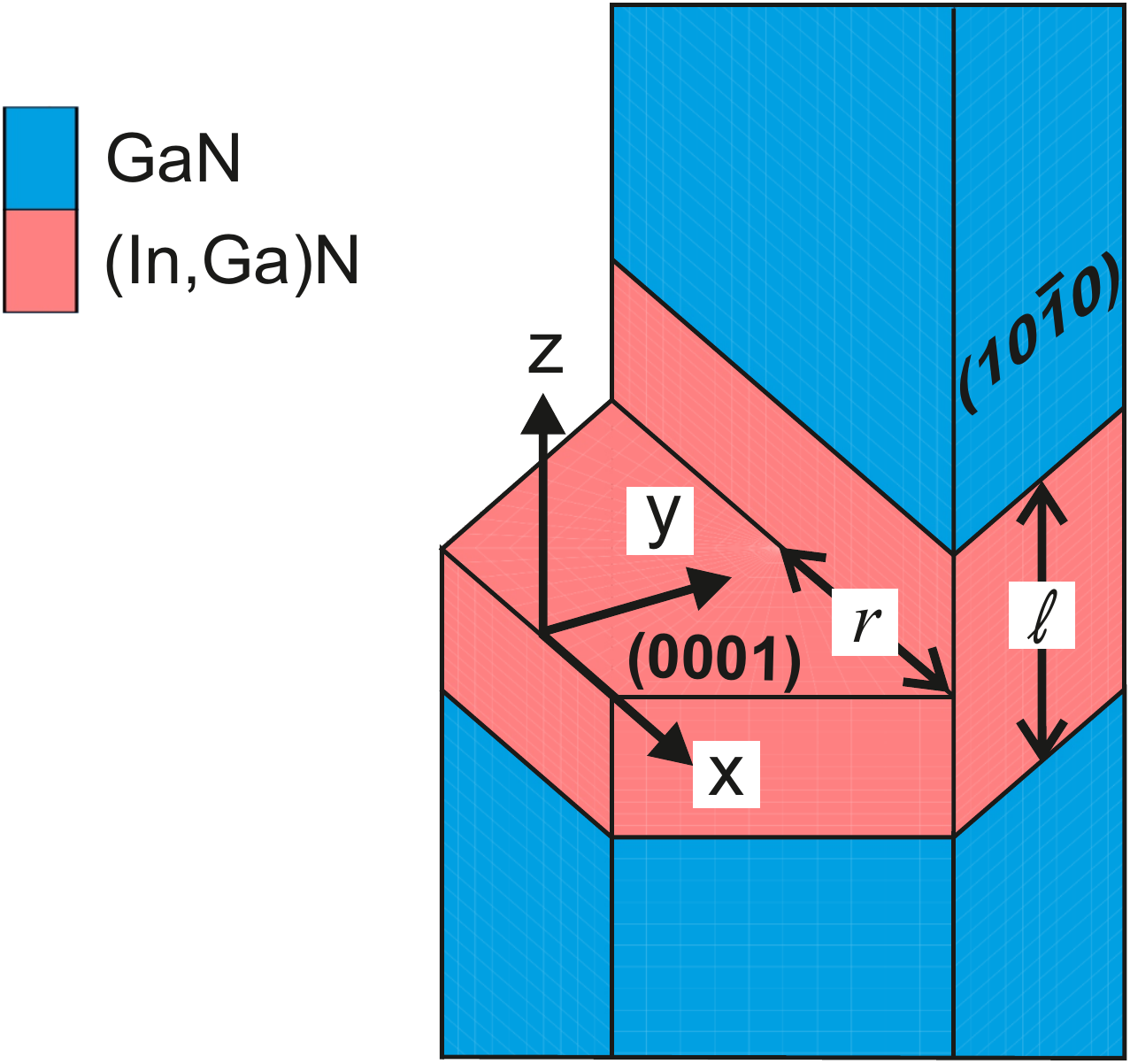}
\caption{Sketch of a hexagonally-shaped GaN nanowire of radius $r$ with an embedded axial (In,Ga)N segment of length $\ell$.}
\label{fig:Model}
\end{figure}

The strain field induced by a lattice-mismatched segment in a NW can be obtained in closed form only when treating the NW as an infinitely long cylinder.\cite{Kaganer2012} To take into account the actual hexagonal cross-sectional shape of the GaN NWs under consideration, we employ the finite element method (FEM) as implemented in the commercial package MSC Marc\textsuperscript{\textregistered}. Our simulations are performed within the framework of linear elasticity theory for which the components of the strain tensor are dimensionless. An important consequence of this fact is that our results can be scaled arbitrarily to smaller or larger dimensions as long as the strain relaxation is purely elastic. In other words, the strain field in the NW will be identical for all axial (In,Ga)N/GaN NW heterostructures with the same ratio of NW diameter and segment length. Our simulations take into account the full elastic anisotropy of the hexagonal group-III nitrides. The elastic constants for GaN and InN are taken from Ref.~\onlinecite{Shimada2006}, and the elastic constants for the ternary (In,Ga)N alloy are obtained by linear interpolation.

\begin{figure*}
\centering
\includegraphics[width=1.0\textwidth]{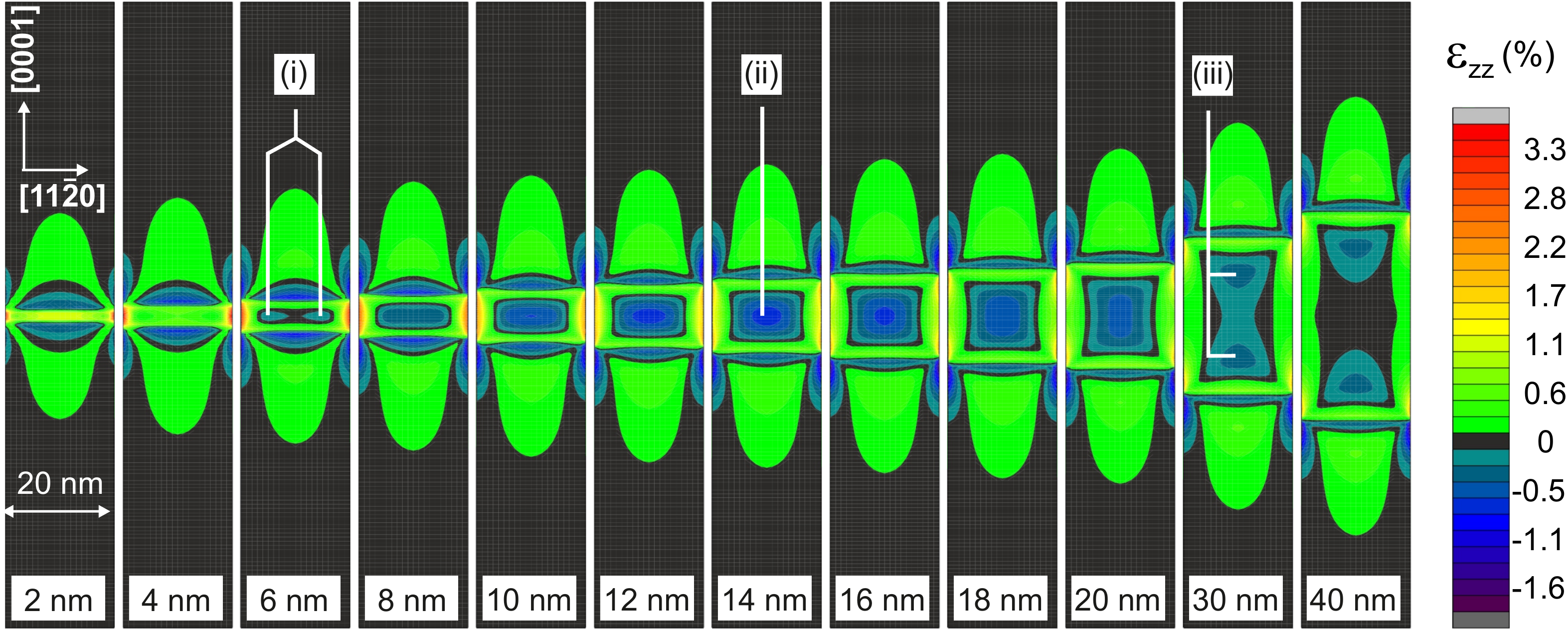}
\caption{Map of the elastic strain component \ezz for a vertical slice across the center of the NW and for different lengths $\ell$ of the embedded \ingan segment from 2 to 40~nm. Starting from a compressively strained \ingan QD with positive \ezz at $\ell = 2$~nm, \ezz gradually decreases and eventually turns negative in a torus-shaped region for the QD with $\ell\approxeq 6$\,nm [(i)]. This region expands with increasing $\ell$ until \ezz assumes a minimum value of about $-0.8$~\% in the center of the \ingan segment with $\ell = 14$~nm [(ii)]. A further increase of $\ell$ results in complete elastic relaxation (\ezz $\to$ 0) in the middle part of the segment with $\ell = 40$~nm [(iii)].}
\label{fig:FEM_h/r}
\end{figure*}

In all what follows, we consider a hexagonally shaped, $\langle0001\rangle$-oriented GaN NW with an embedded In$_{x}$Ga$_{1-x}$N segment with $x = 0.3$ as schematically depicted in Fig.~\ref{fig:Model}. The alloy is treated as a perfectly homogeneous material, i.\,e., the random compositional fluctuations in this material are ignored. For the sake of explicitness, but without loss of generality, we set the NW radius to $r = 10$\,nm. Energy balance considerations for an elastically isotropic material predict that the \ingan segment remains coherently strained for this radius regardless its length $\ell$.\cite{Glas2006} Experimental investigations were focused on very short (In,Ga)N segments which in the following we refer to as quantum disks (QDs) with a thickness of typically not more than 2 to 5~nm.\cite{Guo2011,Wolz2012,Tourbot2013} For an \ingan QD with this thickness, the emission wavelength is expected to be between 600 and 700~nm, making these structures interesting for applications for red light emitting diodes. Thicker QDs would be expected to emit in the infrared spectral range, but for the corresponding planar structures, strain-induced piezoelectric fields within the QDs would inhibit the emission altogether. For NWs, however, the elastic strain relieve at the NW sidewalls may greatly reduce the magnitude of these fields. For a systematic investigation of the evolution of the strain field within the NW, we thus vary the segment length $\ell$ from 2 to 40~nm while keeping the In content and NW radius constant. 

We first focus on the out-of-plane component of the elastic strain tensor, \ezz, as defined by the relative lattice parameter difference with respect to the unstrained bulk material. For a pseudomorphically strained (In,Ga)N/GaN layered system, the (In,Ga)N layer is under biaxial compressive strain characterized by a positive value of \ezz throughout the (In,Ga)N layer. Intuitively, we would expect a positive value for \ezz for an equivalent layer sequence in a NW as well. Since strain can be elastically relieved at the NW surface, \ezz will be decreased particularly close to the surface, and should eventually approach zero in the limit of very thin NWs except for the interfacial boundaries. However, in the following we will demonstrate that the strain relaxation in axial (In,Ga)N/GaN NWs does not proceed in this simple monotonic fashion as intuitively expected.

To follow the evolution of the strain field with increasing $\ell/r$ ratio, we calculate a map of \ezz for different lengths of the embedded \ingan segment as shown in Fig.~\ref{fig:FEM_h/r}. For the 2\,nm thick QD, a strain state similar to an equivalent planar system is established, i.\,e, the QD is under essentially uniform strain with $\upvarepsilon_\mathrm{zz} > 0$ throughout its volume. Strain relaxation mainly occurs in the direct vicinity of the free sidewall surfaces and the \ingagan interfaces, manifesting itself in long-range distortions in the adjacent GaN segments which change their character from tensile ($\upvarepsilon_\mathrm{zz} < 0$) to compressive ($\upvarepsilon_\mathrm{zz} > 0$) with increasing distance from the QD. Furthermore, \ezz attains values of more than 3\% directly at the edges of the hexagonal \ingan QD.

For thicker QDs, this behavior changes profoundly, in that the relaxation affects progressively more of the volume of the \ingan QD itself. For $\ell = 6$\,nm [see panel (i) in Fig.~\ref{fig:FEM_h/r}], a part of the inner core of the QD exhibits the bulk lattice constant of \ingan with $\upvarepsilon_\mathrm{zz} = 0$. Embedded into this region are two radially separated regions with a negative value of \ezz. These regions represent a cut through a torus-shaped strain distribution with maximum compression.

When increasing $\ell$, the spatially separated compressed regions extend toward the center of the segment and eventually merge. As a result, the entire inner core of the \ingan segment is under tensile strain with $\upvarepsilon_\mathrm{zz} < 0$. The maximum magnitude of this tensile strain is reached for $\ell/r\approx 3/2$ [see panel (ii) of Fig.~\ref{fig:FEM_h/r}], for which \ezz approaches a negative value as large as $-0.8$\%. The maximum values for \ezz of about 2\% are again observed at the edges of the \ingan segment.

A further increase of the $\ell/r$ ratio beyond the value of 3/2 reverses the continuous decrease of \ezz in the center of the segment. However, the inner core of the \ingan segment persists to be under tensile strain until the $\ell/r$ ratio attains values larger than 3 [see panel (iii) in Fig.~\ref{fig:FEM_h/r}]. In this case, the minimum in the center of the segment starts to divide into two minima forming again a torus-shaped strain distribution, but now in axial direction. At the same time, the value of \ezz in the center of the segment slowly approaches zero. Finally, a strain state akin to complete elastic relaxation is observed for the middle part of the 40\,nm long \ingan segment. However, the strain field stays complex at the interface to the adjacent GaN segments.

Note that the complete elastic relaxation observed for long segments is not to be confused with the vanishing \ezz for $\ell/r =3/2$ [cf.\ panel (i) of Fig.~\ref{fig:FEM_h/r}]. While the former effect is a simple manifestation of St.\ Venant's principle \cite{love_44,timochenko_51,housner_66} the latter one is as unexpected as it is difficult to understand on an intuitive basis.

\begin{figure}
\includegraphics[width=1.0\columnwidth]{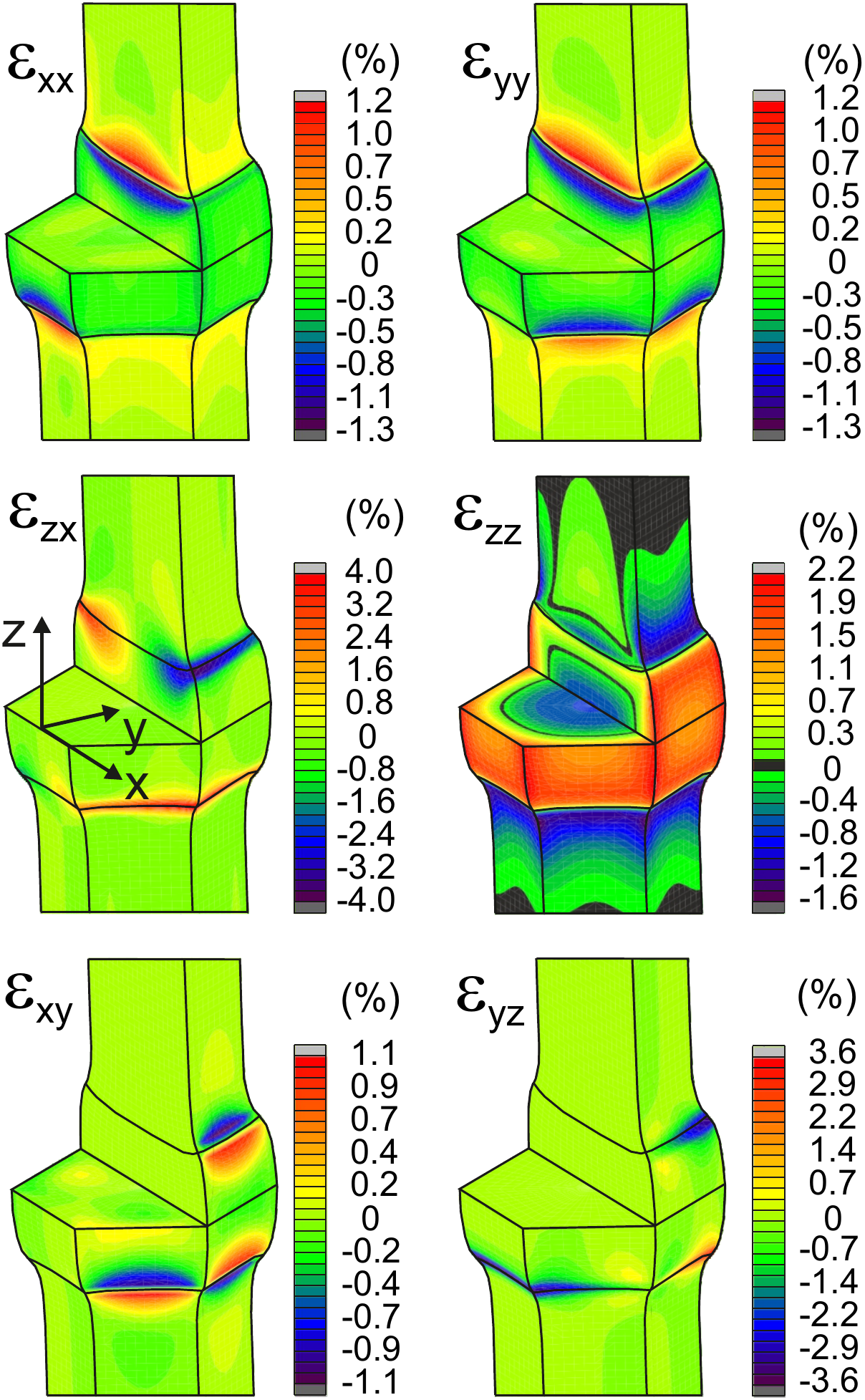} \caption{On and off-axis strain components for the \ingagan NW with $\ell/r = 1$. The in-plane normal components \exx and \eyy suggest a nearly complete strain relaxation, but the out-of-plane normal component \ezz is negative in the inner core of the segment. The strain release at the surface induces a deformation of the \ingan segment, exaggerated here by a factor of 10 for better visibility. This deformation is accompanied by significant shear strain components with out-of-plane contribution, concentrated mainly at the \ingagan interfaces.}
\label{fig:StrainComp}
\end{figure}

To shed light onto this anomaly in the dependence of the out-of-plane strain on the $\ell/r$ ratio, we next consider all components of the strain tensor, and in particular the shear strains. As a representative example, Fig.~\ref{fig:StrainComp} shows these components for the \ingagan NW with $\ell/r = 1$, for which a significant portion of the \ingan segment exhibits a tensile out-of-plane strain (cf.\ Fig.~\ref{fig:FEM_h/r}). Despite of this fact, the in-plane normal components \exx and \eyy are seen to be spatially uniform and small ($-0.3$\%) essentially within the entire \ingan segment. Deviations from this quasi-relaxed in-plane strain state are observed only at the interfaces, where the adjacent materials experience a strong compressive and tensile strain to accommodate the lattice mismatch and to establish a commensurate boundary.

In contrast, the out-of-plane component \ezz is negative in the central part of the segment and continuously increases toward both the free surface and the interfaces to the adjacent GaN segments. On each \{10\={1}0\} facet of the segment surface, \ezz is smallest at the center of the facet and increases toward the hexagonal edges as well as toward the interfaces with a maximum value of about 2\%. This variation of \ezz is accompanied by a significant convex deformation of the shape of the segment at the free sidewalls.

This deformation of the \ingan segment necessitates non-zero shear strains. In fact, all three shear-strain components are significant, and particularly so those with out-of-plane components (\ezx and \eyz), which reach magnitudes as large as 4\%. These shear components allow an almost complete in-plane relaxation of the \ingan segment via the convex deformation of the NW sidewalls. It is this deformation which results in an axial contraction, and is thus directly responsible for the negative out-of-plane strain observed in this study. Additionally, due to the small Poisson ratio of GaN the \ingan segment induces a volume change of 1:0.969 deviating from an equivalent layered structure which shows a volume change close to one.

To access the local strain state of a single NW experimentally is a difficult task. A particularly powerful technique to investigate the three-dimensional strain distribution within single (In,Ga)N/GaN NWs is nanofocus x-ray diffractometry\cite{Robinson2009,Stangl2014,Keplinger2015}. This non-destructive technique combines a high angular resolution in reciprocal space with a sub-\textmu m spatial resolution\cite{Hanke2008,Gulden2011,Dubslaff2011}. To assess the feasibility to experimentally measure the strain distribution for single (In,Ga)N/GaN NWs, we simulate the impact of a strained (In,Ga)N segment in a GaN NW on the diffuse x-ray scattering pattern of this NW heterostructure. Using the kinematical expression

\begin{align}
I(\bm{q})=\abs{\sum_i f_i(\bm q) e^{\ii {\bm q}\cdot[{\bm r}_i+{\bm u}({\bm r}_i)]}}^2
\end{align}
\noindent
with the reciprocal space vector {$\bm q$} = (\qx, \qy, \qz), the form factors $f_i(\bm q)$, and the atom coordinates ${\bm r}_i$ , we compute the three-dimensional intensity pattern around the Bragg reflection in reciprocal space. The atom coordinates are interpolated using the displacement field $\bm u$ obtained from FEM giving the displacement of the atom positions ${\bm u}({\bm r}_i)$. As an example, we will discuss four simulated reciprocal space maps (RSM), showing the diffusely scattered intensity around the symmetric GaN(0004) reflection. We assume a spot size of 120$\times$120~nm$^2$ and the strain fields are taken from the series of Fig.~\ref{fig:FEM_h/r}.

\begin{figure}
\includegraphics[width=1.0\columnwidth]{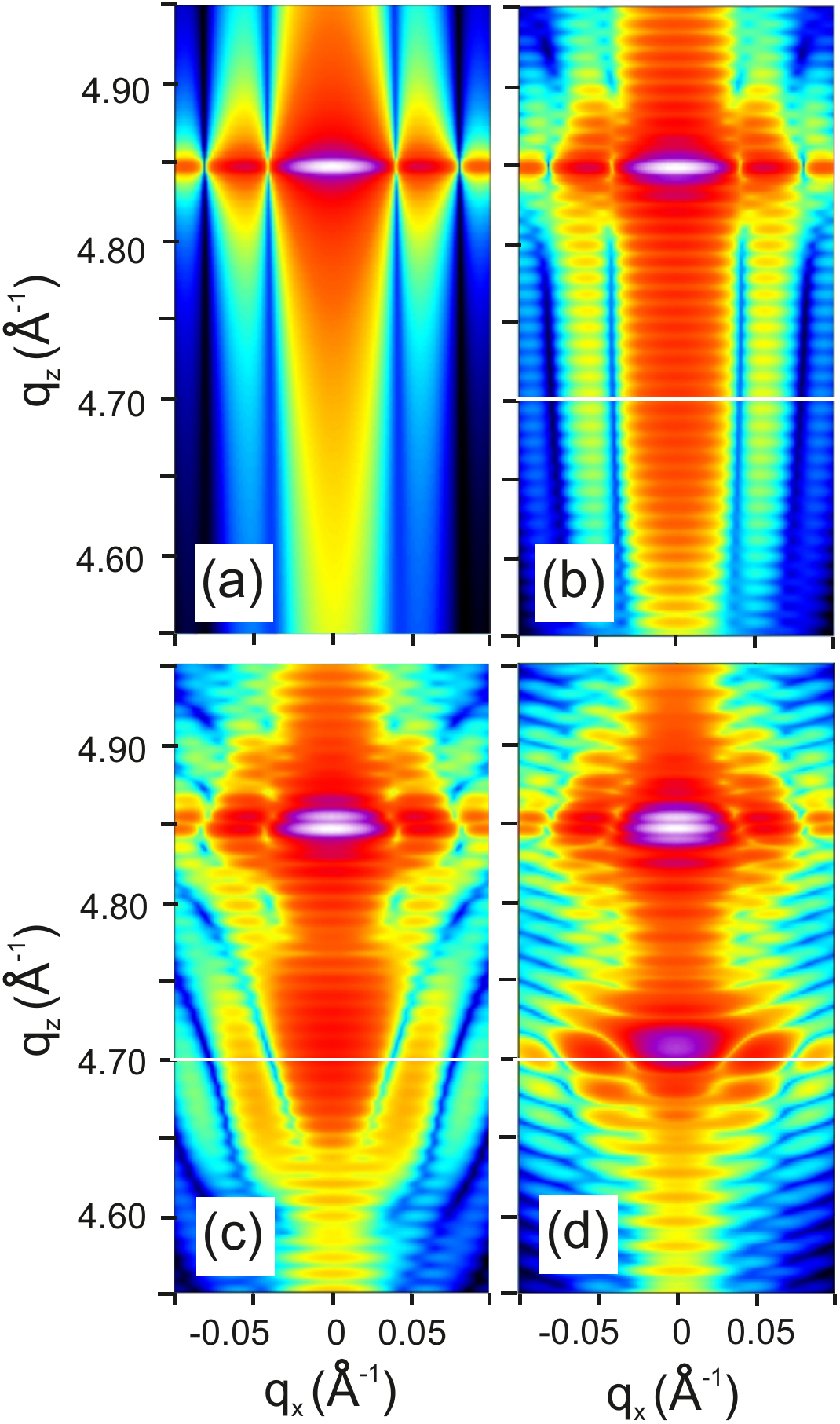}
\caption{Reciprocal space maps close to the symmetric GaN(0004) reflection calculated in the frame of the kinematical scattering theory. (a): RSM for a bare GaN NW with radius $r = 10$\,nm. (b), (c) and (d): RSMs  for GaN NWs with embedded (In,Ga)N segments of length $\ell= 4$ (b), 10 (c) and 40~nm (d). The white line indicates the expected peak position for relaxed \ingan.}
\label{fig:ScatSim}
\end{figure}

In Fig.~\ref{fig:ScatSim}(a), we consider a bare GaN NW yielding a peak at \qz=~4.85~\AA$^{-1}$ which is the expected position for bulk GaN. This peak is modulated by finite size oscillations due to the NW geometry. The pattern changes as soon as strain is introduced into the NW due to the presence of an \ingan segment. The RSM shown in Fig.~\ref{fig:ScatSim}(b) for a 4\,nm thick QD exhibits no distinct peak related to \ingan, but the phase shift induced by the QD results in pronounced interference fringes. For $\ell = 10$\,nm, the FEM simulation predicts negative elastic strain within the QD volume, which in turn will shift the peak position of relaxed \ingan toward larger values of \qz as indeed seen in Fig.~\ref{fig:ScatSim}(c). Besides the change in the position of the Bragg peak, the diffuse part at $q_\mathrm{x} \neq 0$ serves as a sensitive fingerprint of the established strain field within a NW. Due to the comparatively large volume of relaxed \ingan, the RSM of the 40\,nm long segment exhibits a clear peak with high intensity close to the position of relaxed \ingan indicated by the white line in Fig.~\ref{fig:ScatSim}(d). The complex strain field at the interfaces manifests itself by the strong modulations around this peak. Based on the simulated RSMs, we conclude that an \ingan segment embedded in a GaN NW has a significant, experimentally accessible impact on the diffraction pattern, and nanofocus x-ray diffraction experiments are thus suitable to probe the local strain distribution in these NW heterostructures.

The results obtained in this work for $\langle0001\rangle$-oriented (In,Ga)N/GaN NWs are also valid for other axial semiconductor NW heterostructures, including all $\langle0001\rangle$-oriented wurtzite NWs, but also $\langle111\rangle$-oriented NWs composed of diamond and zincblende materials such as Ge/Si\cite{Hanke2007,Swadener2009,Wen2015}, Si/GaP\cite{Hocevar2012}, InAs/InP\cite{Haapamaki2011,Gotoh2015}, CdTe/ZnTe\cite{DLUZEWSKI2010}. For all these materials systems, the strain anomaly investigated in the present work is not only of academic interest, but has potentially far-reaching consequences for their application in electronic and optoelectronic devices. In fact, our study shows that the length-to-radius ratio can be used to engineer the strain field and in particular to tune the out-of-plane strain component from compressive to tensile. We can use this new degree of freedom to manipulate the band gap, the effective mass, and the magnitude and direction of the piezoelectric fields in the strained insertion.

The authors are indebted to Oliver Marquardt for valuable discussions and a critical reading of the manuscript. Special thanks are due to Vladimir Kaganer for performing analytical calculations which confirm the anomalous strain state we have discussed in the present work. Financial support by the German Research Foundation DFG (project HA3495/9-1) is gratefully acknowledged.

\end{document}